\documentclass[12pt,preprint]{aastex}

\usepackage{graphicx}

\slugcomment{}

\shorttitle{Activity in Phaethon}
\shortauthors{Jewitt and Li}

\begin{document}

\title{Activity in Geminid Parent (3200) Phaethon}

\author{David Jewitt$^{1}$ and Jing Li$^2$, }
\affil{$1$ Dept. Earth and Space Sciences, Institute for Geophysics and Planetary Physics and Dept. Physics and Astronomy, UCLA \\
$2$ Institute for Geophysics and Planetary Physics, UCLA \\
}
\email{jewitt@ucla.edu, jli@igpp.ucla.edu}

\begin{abstract} 
The asteroid (3200) Phaethon is widely recognized as the parent of the Geminid meteoroid stream. However, it has never shown evidence for on-going mass loss or for any form of comet-like activity that would indicate the continued replenishment of the stream.  Following an alert by Battams and Watson (2009), we used NASA's STEREO-A spacecraft to image Phaethon near perihelion, in the period UT 2009 June 17 - 22, when the heliocentric distance was near 0.14 AU.  The resulting photometry shows an unexpected brightening, by a factor of two, starting UT 2009 June 20.2$\pm$0.2, which we interpret as an impulsive release of dust particles from Phaethon. If the density is near 2500 kg m$^{-3}$, then the emitted dust particles must have a combined mass of $\sim$2.5$\times$10$^8 a_1$ kg, where $a_1$ is the particle radius in millimeters. Assuming $a_1$ = 1, this is approximately 10$^{-4}$ of the Geminid stream mass and to replenish the stream in steady-state within its estimated $\sim$10$^3$ yr lifetime would require $\sim$10 events like the one observed, per orbit.  Alternatively, on-going mass loss may be unrelated to the event which produced the Phaethon-Geminid complex.  An impact origin of the dust is highly unlikely. Phaethon is too hot for water ice to survive, rendering unlikely the possibility that dust is ejected through gas-drag from sublimated ice.  Instead, we suggest that Phaethon is essentially a rock comet, in which the small perihelion distance leads both to the production of dust (through thermal fracture and decomposition-cracking of hydrated minerals) and to its ejection into interplanetary space (through radiation pressure sweeping and other effects).
 
\end{abstract}


\section{Introduction}
Near-Earth asteroid 3200 Phaethon (formerly 1983 TB) is dynamically associated with (and presumed to be the parent of) the Geminid meteoroid stream (Whipple 1983, Fox et al. 1984, Gustafson 1989, Williams and Wu 1993). However, whereas most established meteoroid stream parents are clearly cometary in nature (e.g. Jenniskens 2008), the orbit of Phaethon is that of an asteroid.  Its semimajor axis, eccentricity and orbital inclination are, respectively, $a$ = 1.271 AU, $e$ = 0.890 and $i$  = 22.2$\degr$, and the Tisserand parameter with respect to Jupiter is $T_J$ = 4.5 (whereas the conventional comets have $T_J <$ 3; Kresak 1980).   The perihelion distance is a remarkably small $q$ = 0.14 AU.  Phaethon is roughly 5 km in diameter with $V$-band geometric albedo $\sim$0.17 (Veeder et al. 1984; a somewhat larger diameter of 7.0$\pm$0.5 km and smaller albedo, $p_v$ = 0.05$\pm$0.01, are reported by Kraemer et al. 2005, but their measured thermal emission spectrum (their Figure 20) is unconvincing).  No evidence for mass loss from Phaethon has ever been reported, either in gas (Cochran and Barker 1984, Chamberlin et al. 1996, Wiegert et al. 2008) or dust (Hsieh and Jewitt 2005).   

Recently, other objects have been linked dynamically to Phaethon and therefore to the Geminids (Ohtsuka et al. 2006, 2008).    The $\sim$1.3 km diameter asteroid 2005 UD shares both a dynamical link to Phaethon, and a physical link in the sense that both it and Phaethon are optically blue.  Since blue asteroids constitute only about 4\% of the general asteroid population, the finding that 2005 UD and Phaethon have related orbits and are both blue is highly suggestive of a connection or common origin (Jewitt and Hsieh 2006, Kinoshita et al. 2007).  Asteroid 1999 YC may also be dynamically related to Phaethon but its surface appears more nearly neutral with respect to the color of the Sun (Kasuga and Jewitt 2008).   This ensemble of small bodies, which probably extends to smaller objects yet to be discovered, is sometimes referred to as the Phaethon-Geminid Complex (PGC; Ohtsuka et al. 2006). It is presumably the product of disintegration of a precursor object but the mechanism behind the disintegration, and whether these are products of a catastrophic event or a continuing process, are both unknown (Kasuga 2009).  The age of the Geminid stream is uncertain but it is probably young.  Numerical models including the effects of radiation forces and planetary gravitational perturbations give ages in the 600 yr to 2000 yr range (Gustafson 1989, Williams and Wu 1993, Ryabova 2007).  The PGC break-up probably pre-dates the stream formation, but this event is not observationally constrained.

The mass of Phaethon, represented as a 5 km diameter sphere of assumed density $\rho$ = 2500 kg m$^{-3}$ (although estimates are model-dependent and scattered over a factor of two, this is the ``typical'' density of  Geminid meteors according to Borovicka et al. 2009), is $M_{3200}$ = 1.6$\times$10$^{14}$ kg.  The mass of the Geminid meteoroid stream is highly uncertain, with published estimates in the range $M_s$ = 10$^{12}$ to 10$^{13}$ kg (Hughes and McBride 1989, Jenniskens 1994).  The mass in the stream thus corresponds to a surface shell on Phaethon of order 5 to 50 m thick.  Questions of central importance relate to  how matter from Phaethon was ejected and when the ejection occurred.  

Watson (reported in Battams and Watson 2009) observed that Phaethon brightened  by about 2 magnitudes or more a few hours after  perihelion (UT 2009 June 20.301) in data from the SECCHI HI-1A (STEREO) mission.  They also reported that Phaethon showed a non-stellar morphology and suggested that an unspecified interaction with the solar wind might be the cause.  These observations are particularly interesting because, if real, they would constitute the first clear examples of on-going mass-loss from Phaethon.  In this paper, we examine in detail the near-perihelion data on Phaethon from the STEREO mission with the following questions in mind; 1) is the reported perihelion brightening real? and, if so, 2)  what are its properties? and 3) what is its cause?

\section{Observations}
NASA's STEREO mission consists of two spacecraft in Earth-like orbits but which drift relative to Earth in opposite directions around the Sun (Kaiser et al. 2008).  Cameras on each spacecraft view the sun from different vantage points and together provide a stereoscopic perspective on solar coronal structures.  STEREO A (ahead of the Earth) and STEREO B (behind the Earth) carry identical instruments, including the Heliographic Imagers, containing two cameras called HI-1 and HI-2 respectively. These have 20 (HI-1) and 70 (HI-2) degree-wide fields of view with their centers pointed 14.0 and 53.7 degrees from the Sun along the ecliptic plane (Eyles et al. 2009).  The charge-coupled device (CCD) HI detectors are 2048$\times$2048 pixels in size but are binned 2$\times$2 on-board, giving angular pixel scales in the resulting images of 70 arcsec (HI-1) and 250 arcsec (HI-2).  For the Phaethon measurements, we used the HI-1 camera on STEREO A, exclusively.  For these cameras, the CCD quantum efficiency peaks near 93\% at 5500\AA~wavelength, falling to $\sim$20\% at 3000\AA~in the blue and 9400\AA~in the red.  However, the effective passband  is limited by the transmission of the camera optics, this being a flat-topped function centered near 6700\AA~and with full-width at half-maximum 1400\AA~(see Figure 27 of Eyles et al. 2009).  This passband is close to the classical astronomical photometric ``R'' band and, for the purposes of this paper, we assume that they are the same.

Each image from the HI-1 camera consists of 30 separate images, of 40 seconds integration, for a combined integration time of 1200 seconds.  Because of instrumental overheads, these 30 constituent images are accumulated over a period of 1800 seconds. New combined images are produced every 2400 seconds, meaning that  the duty cycle (defined as integration time divided by total elapsed time) for the HI-1 camera is 50\% (Eyles et al. 2009).  

\begin{figure}[]
%\epsscale{1.0}
%\label{rrr}
\begin{center}
%%\plotfiddle{f101.ps}{0.5cm}{-90.}{.4}{.4}{-10000}{.00}
%\plotone{figure6.pdf}
%\plotone{figure1.pdf}
%\plotone{jewitt_f1.pdf}
\includegraphics[width=1.0\textwidth]{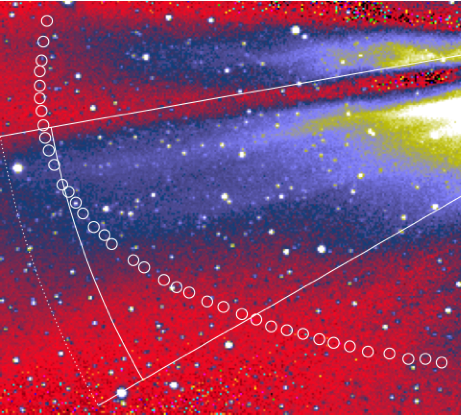}
\caption{Trajectory of 3200 Phaethon in the field of STEREO camera HI-1A.  The coronal image was taken at UT 2009 June 20 08h 49m.  Solid arc has radius equal to Phaethon's 0.14 AU perihelion distance and the segment angle of 20$\degr$ is chosen to highlight the streamer rooted at position angle 110$\degr$.  Ecliptic North is to the top and East to the left of this 5.0$\times$4.5  degree wide portion of the field of view.   The Sun is off to the right. \label{path}} 
\end{center} 
\end{figure}

We identified Phaethon in HI-1 on STEREO A (HI-1A) by its position and motion relative to the fixed stars in the 2009 June 17.5  (day-of-year DOY 168.5) to 2009 June 22.2 (DOY 173.2) period (Figure \ref{path}).  The solar elongation increased from 5 deg to 8.5 deg in this interval.  On earlier dates, Phaethon fell close to (or beyond) the edge of the field of view and could not be detected. On later dates, Phaethon was well-placed on HI-1A but was too faint to be usefully measured.   Of 200 images taken between June 17 and 22, Phaethon was detected in 165 images but some were compromised by nearby field stars and other sources of contamination.  Data from the HI-1B camera (on the other STEREO spacecraft) were also examined but Phaethon could not be identified there (opposite to the finding of Battams and Watson 2009), probably as a result of the larger phase angles (and greater phase dimming) and smaller elongation angle as seen from this spacecraft.  

Visual examination of the data in the form of a movie showed an apparent and dramatic brightening of Phaethon, qualitatively consistent with the report of Battams and Watson (2009).  However, the  coronal background changes in brightness and morphology, raising the possibility that the visual brightening of Phaethon might be an artifact of the spatially and temporally variable and complex field.    Figure (\ref{path}) plots the path of Phaethon from UT June 19.506 to June 21.173.  The bright coronal structure is marked by a cone having apex angle 20$\degr$ and position angle of 110$\degr$.   We searched the SOHO Coronal Mass Ejection Catalog for coronal mass ejection events (CMEs) observed by the Large Angle and Spectrometric Coronograph (LASCO) near this position angle, but without success.
 Instead, the coronal brightening appears to be a large-scale quiescent coronal streamer (Gosling et al. 1981, Raymond et al. 1997).   Supporting this conclusion is the observation that CMEs are energetic events which typically last for a few hours while the enhancement in Figure (\ref{path}) persisted for $\sim$1.5 days.  

%Software onboard the spacecraft removes cosmic rays by searching for large ($>$5$\sigma$ significance) differences in the counts in each pixel between consecutive images.   We checked that this algorithm, which is integral to the data pipeline and over which we have no control, does not falsely flag and remove Phaethon as a cosmic ray because its motion in the 60 seconds between consecutive images is too slow to carry it out of a pixel. 

\subsection{Photometry with STEREO}

To quantify the apparent brightening in Phaethon, we obtained photometric measurements.  For our purposes, coronal structures in the STEREO images represent an unfortunate, time-dependent distraction.  As a first step, we computed the median of all images taken in the above period and subtracted it from the individual images in the sequence.  The resulting images show Phaethon and the field stars superimposed on a much-reduced background consisting of coronal structures whose position and/or surface brightness changed appreciably during the 5 days of measurement.  

We used aperture photometry to measure the data, paying particular attention to the selection of the optimum aperture size in view of the large pixels and the structured coronal background.  Small apertures exclude a significant fraction of the light from astronomical images while large apertures suffer from excessive uncertainties due to noise and gradients in the sky background.  The optimum aperture size is, in part, a compromise between these effects.  Very small apertures cannot be used, even though the pixel size is very large (70 arcseconds), because light from a point source occupies more than one pixel. Moreover, the distribution of light amongst pixels varies from image to image as the center of light moves with respect to the pixel grid.  

Another factor limiting the use of very small apertures is  image trailing, caused by the motion of the HI-1 field of view as it follows the Sun at a fixed solar elongation of 14 degrees, and also by the Keplerian motion of Phaethon.  The motion of the spacecraft tracks the CCD at about 150 arcsec hr$^{-1}$ East and 50 arcsec hr$^{-1}$ North.  The motion of Phaethon relative to the fixed stars varied from 390 to 180 arcsec hr$^{-1}$ East and 100 arcsec hr$^{-1}$ South to 150 arcsec hr$^{-1}$ North, in the Jun 17 - 22 period.  In the 1800 seconds required to accumulate a single HI-1 camera image, Phaethon moved relative to the CCD by at most 120 arcseconds (1.7 pixels) in RA and 100 arcseconds (1.4 pixels) in declination.  By experimentation, we selected a photometry box 3$\times$3 pixels (210$\times$210 arcseconds) in size for all objects.  Sky subtraction was obtained using the median of the 16 contiguous pixels within a box having dimensions 5$\times$5 pixels (350$\times$350 arcseconds). The use of the median provides some protection from noise fluctuations and from sky pixels contaminated by faint field stars, a significant problem given the large pixel scale.  We rejected images in which visual inspection revealed the close proximity of a field star brighter than Phaethon. Sky boxes larger than those employed were found inappropriate because they tend to sample spatial gradients in the coronal surface brightness and so have reduced relevance to the background brightness in the pixels containing the object.  

\begin{figure}[]
%\epsscale{0.95}
\begin{center}
%%\plotfiddle{f101.ps}{0.5cm}{-90.}{.4}{.4}{-10000}{.00}
%\plotone{figure6.pdf}
%\plotone{stars_lightcurves.pdf}
%\plotone{jewitt_f2.pdf}
\includegraphics[width=1.0\textwidth]{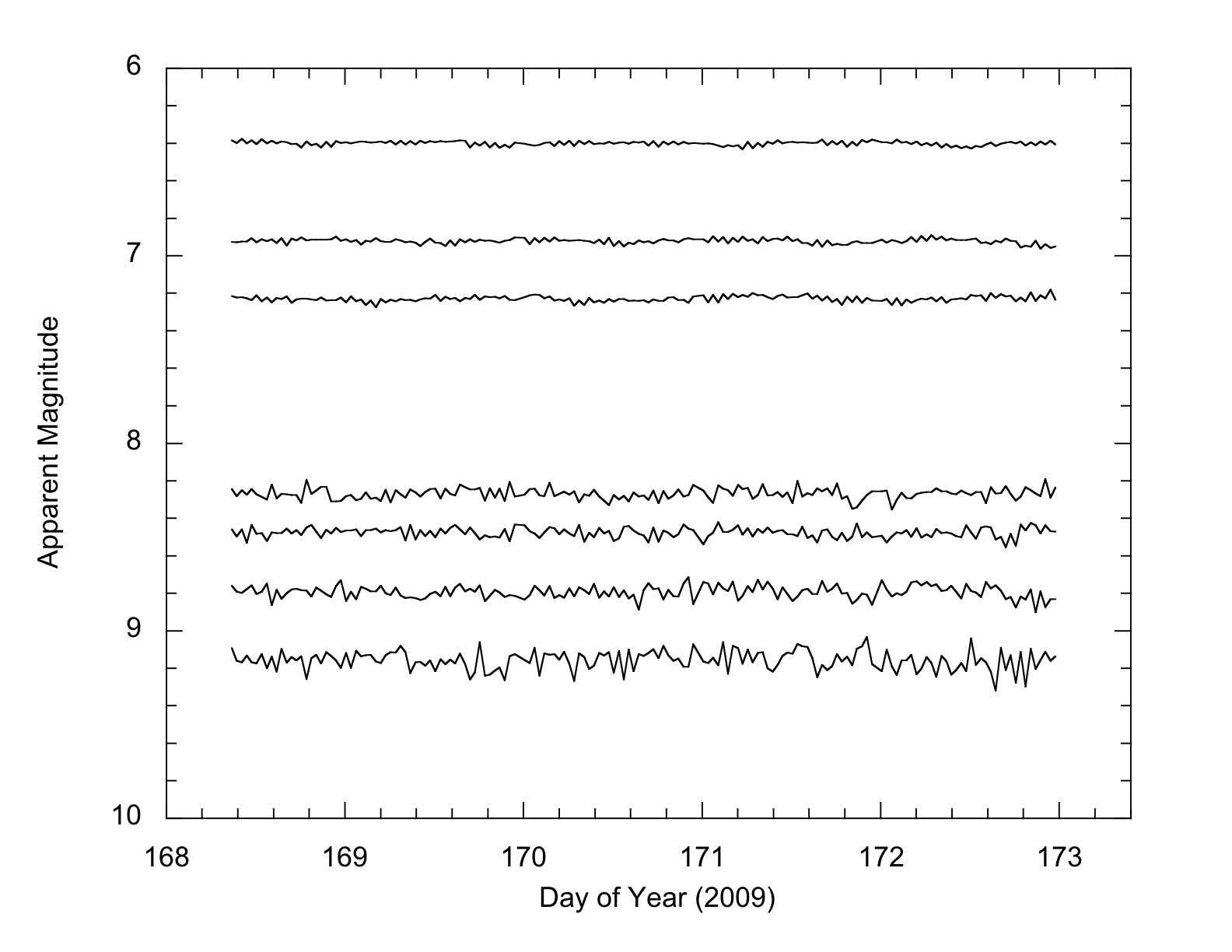}
\caption{Sample photometry of field stars (numbered) extracted from near the path of 3200 Phaethon, plotted as a function of time.  The photometric stability of the system is demonstrated. \label{stars}} 
\end{center} 
\end{figure}

Photometric calibration was achieved using field stars in the images containing Phaethon.  We used measurements of the field stars HD 92184, 91667, 88724, 88680, 87739, 87178 and 86340 (all spectral classes FGK) taken using the same photometry box sizes as used on Phaethon.  The V-magnitudes and V-R colors of these stars were taken from the SIMBAD on-line database and used to convert instrumental magnitudes in the STEREO data to apparent magnitudes in the HI-1 (approximately R) bandpass.  We estimate that the uncertainty of this transformation is about $\pm$0.1 mag.  Repeated measurements of the field stars show that the photometry is precise, with 1$\sigma \sim \pm$0.01 to $\pm$0.03 mag., depending on the brightness of the star (Figure \ref{stars}).

\begin{figure}[]
%\epsscale{0.95}
%\label{bee}
\begin{center}
%%\plotfiddle{f101.ps}{0.5cm}{-90.}{.4}{.4}{-10000}{.00}
%\plotone{figure6.pdf}
%\plotone{counts_vs_sky.pdf}
%\plotone{jewitt_f3.pdf}
\includegraphics[width=1.0\textwidth]{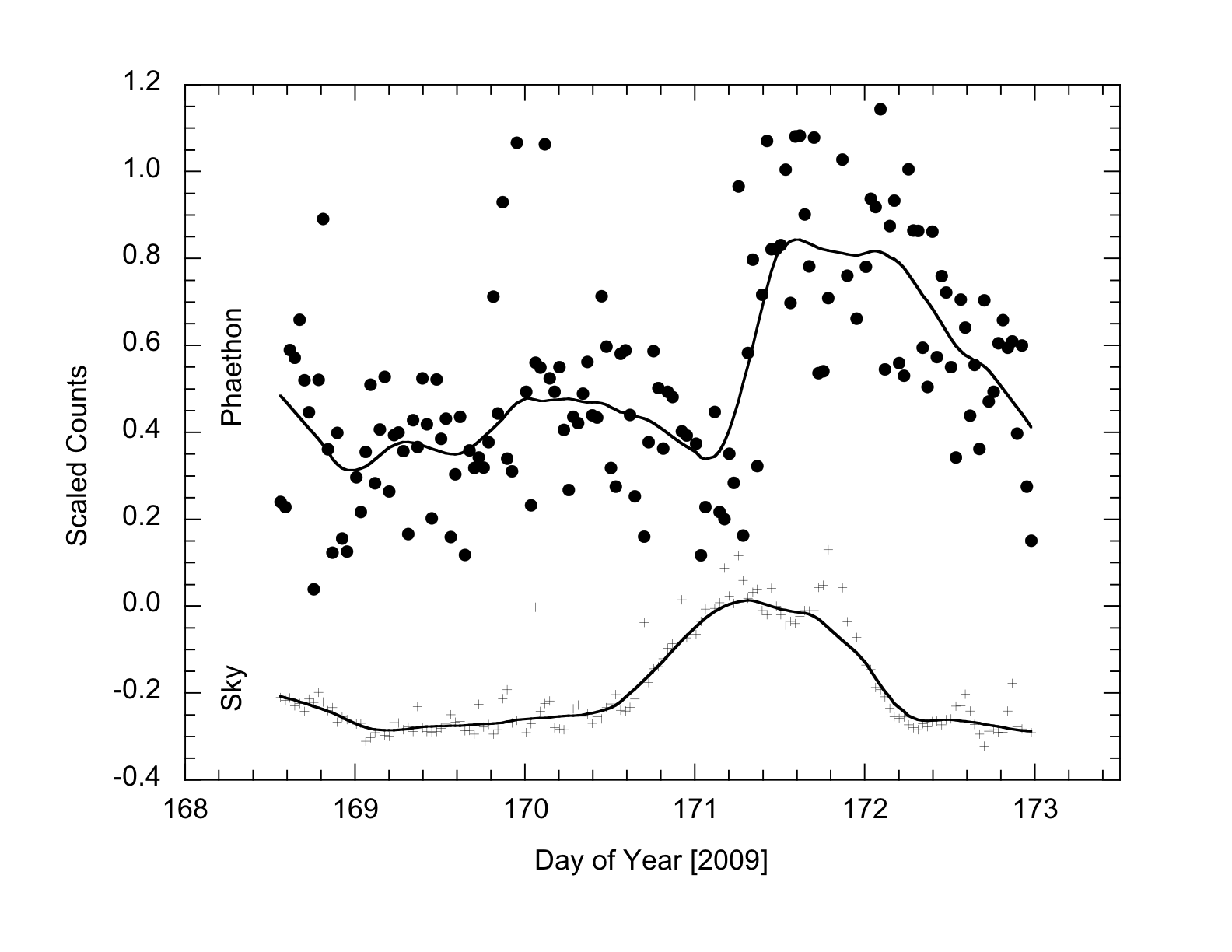}
\caption{Sky-subtracted Phaethon photometry (upper) and the associated sky (lower), versus Day of Year in 2009.  The Phaethon signal is the sum of the 9 pixels within a 3$\times$3 pixel box, with the sky level subtracted from each.  The sky is the median of the 16 pixels contained within nested squares of width 3 pixels and 5 pixels.   To avoid overlap in the plot, 0.35 counts have been subtracted from the sky signal data. \label{sky}} 
\end{center} 
\end{figure}

Figure (\ref{sky}) shows the sky-subtracted Phaethon photometry (black circles) as a function of time.  A running-mean smoothed line has been drawn through the data to guide the eye, and the data are presented as counts (not magnitudes) so that negative excursions in noisy parts of the figure can be shown.  Phaethon's lightcurve is divided into two portions.  For DOY $<$ 171.2$\pm$0.2, the brightness is approximately constant with time.  For DOY $>$ 171.2, Phaethon shows a brightness enhancement, peaking at DOY = 171.8$\pm$0.2 and declining steadily thereafter. Also shown in the figure is the sky background determined at the location of Phaethon.  

The brightening in the sky background in the period  170.6 $< DOY <$ 172.2, with a maximum near DOY = 171.3, is caused by a large scale coronal streamer (see Figure 1 highlighted by the cone) across which Phaethon moved.  While at first we thought that the background might be implicated in the apparent brightening of Phaethon, the background rises $\sim$0.5 day before Phaethon and is already subtracted from the plotted Phaethon photometry; it cannot be the cause of variations seen there.  

\begin{figure}[]
%\epsscale{0.95}
%\label{skycounts}
\begin{center}
%%\plotfiddle{f101.ps}{0.5cm}{-90.}{.4}{.4}{-10000}{.00}
%\plotone{figure6.pdf}
%\plotone{Counts_vs_DOY.pdf}
%\plotone{jewitt_f4.pdf}
\includegraphics[width=1.0\textwidth]{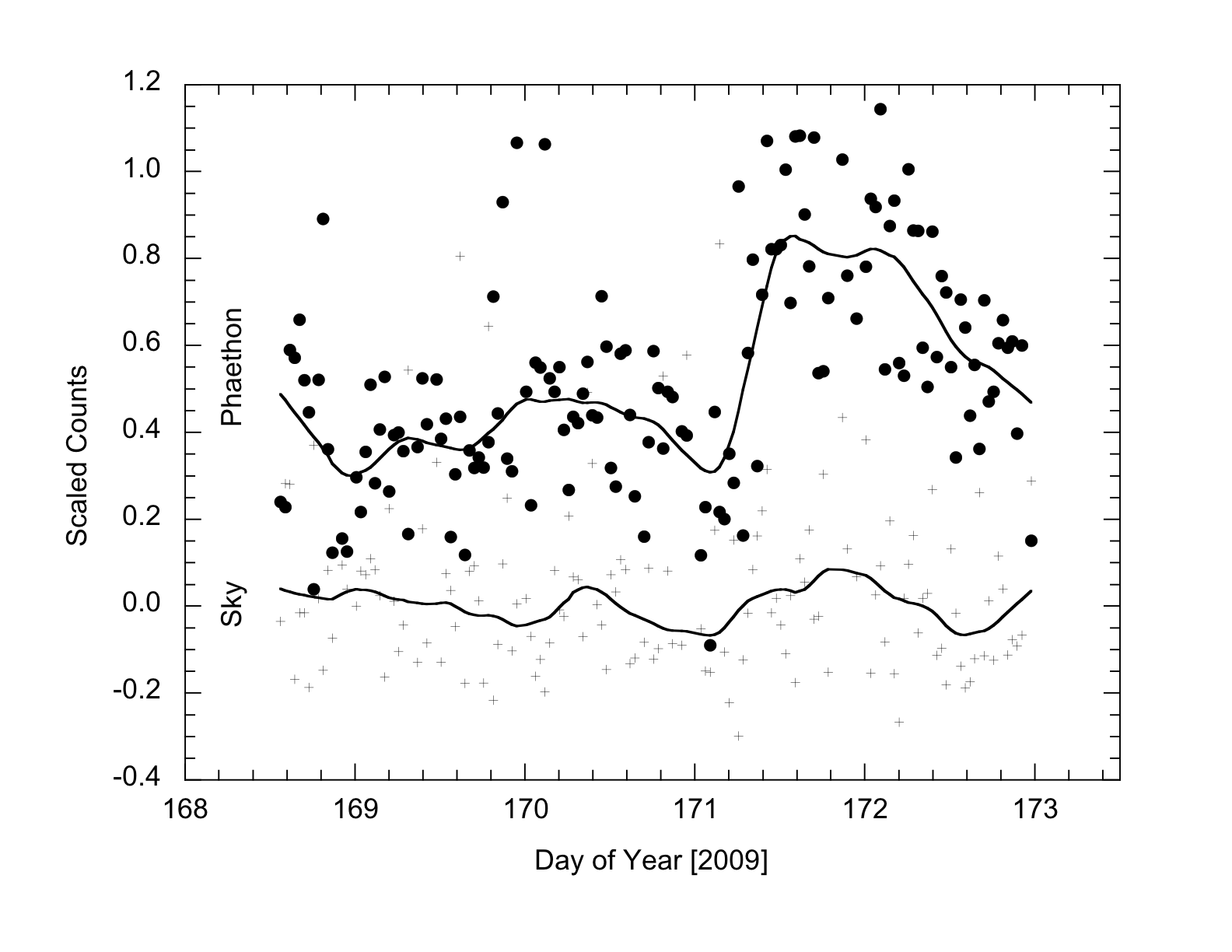}
\caption{Sky-subtracted data number counts for Phaethon and for a sample patch of sky displaced from Phaethon by 5 pixels (350 arcsec) East and 5 pixels South.  The photometry aperture for both Phaethon and the sky was a box of 3$\times$3 pixels while the background was computed from the median of the pixels within a surrounding aperture 5$\times$5 pixels.  \label{lightcurve}} 
\end{center} 
\end{figure}

To further test the accuracy of the background removal procedure, we obtained photometric measurements of blank  sky at several positions following Phaethon but  offset from  it by a fixed amount.  The offsets were  close enough to ensure a common coronal background.  As expected, the measurements of blank sky are consistent with zero, albeit with small fluctuations caused by field stars and residual coronal structures in the CCD image.  One such measurement is shown in Figure (\ref{lightcurve}) (small crosses, again with a smoothed line overplotted) and the lightcurve of Phaethon is included for comparison.  From this and Figure (\ref{sky}), we conclude that Phaethon's brightness increased after DOY = 171.2$\pm$0.2 by a factor of two, and that this increase is not coincident with the coronal brightening but occurred $\sim$0.5 day after it.  From this and other measurements of the sky having different offsets from Phaethon, we are confident both that Phaethon is strongly detected and that large scale variations in the brightness of Phaethon are not caused by imperfect subtraction of the coronal sky background.

\begin{figure}[]
%\epsscale{0.95}
\begin{center}
%%\plotfiddle{f101.ps}{0.5cm}{-90.}{.4}{.4}{-10000}{.00}
%\plotone{figure6.pdf}
%\plotone{sb_profiles.pdf}
%\plotone{jewitt_f5.pdf}
\includegraphics[width=1.0\textwidth]{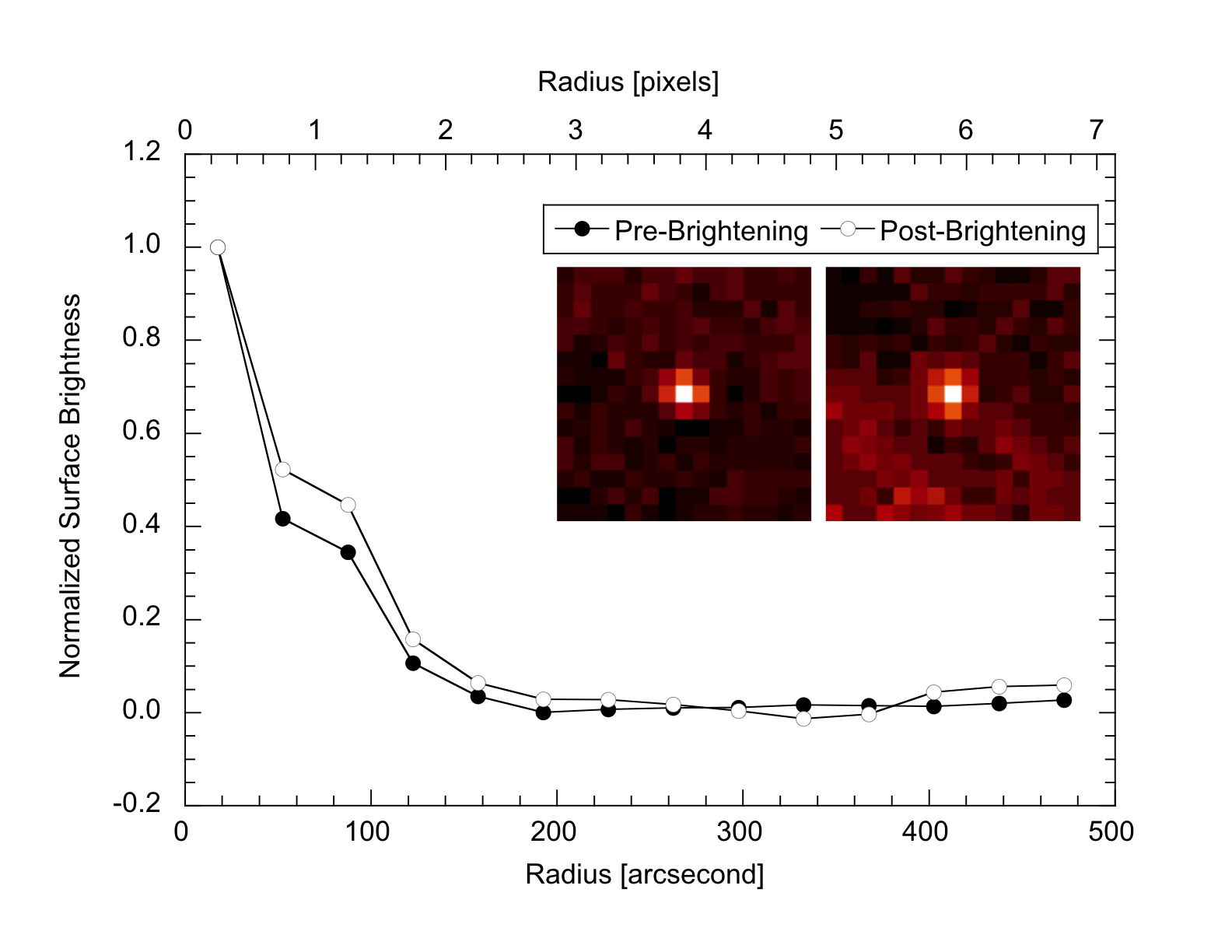}
\caption{Surface brightness profiles of Phaethon  pre-brightening (filled circles) and post-brightening (empty circles), computed as described in the text. The upper and lower axes show the radius in pixels and arcseconds, respectively.  Both profiles have been normalized to unity in the central pixel. Inset figures show (left) the pre-brightening and (right) post-brightening images used to determine the profiles. \label{sb_profiles}} 
\end{center} 
\end{figure}

The surface brightness profiles computed before (76 images in the period UT 2009 June 17.90 to 20.25) and after (55 images in the period UT 2009 June 20.25 to 21.90) brightening are shown in Figure (\ref{sb_profiles}).   Insets in Figure (\ref{sb_profiles}) show the image composites produced by shifting individual images of Phaethon to a common center, subtracting the coronal background and computing the median of the set.  To make the profiles, we measured the data counts within a set of concentric annuli each 0.5 pixel (35 arcsecond) wide and extending out to radius 7 pixels (490 arcseconds), from the image centroid.    The profiles, normalized to unity at the central pixel, are similar in shape, with the post-brightening profile being slightly wider than the pre-brightening profile (Figure \ref{sb_profiles}).  A fit to the profiles gives full-width at half-maximum equal to 1.66 pixel and 1.83 pixel, for the pre- and post-brightening images.  However, the very angular resolution of the data introduces considerable uncertainty into the image profile, depending on the precise location of the image centroid with respect to the pixel grid.  Furthermore, fluctuations in the background affect the profile (the diagonal band in the post-brightening image is residual structure in the scattered light field significantly above local fluctuations due to noise on the sky).  Within these pixellation and background uncertainties, we conclude that the measured profiles are consistent with the point-spread function in the images (i.e. with the profiles of field stars) and that there is no convincing evidence for either a coma or a tail on Phaethon.

\subsection{Scattering function}
\begin{figure}[]
%\epsscale{0.95}
\begin{center}
%\plotone{V_vs_alpha.pdf}
%\plotone{jewitt_f6.pdf}
\includegraphics[width=1.0\textwidth]{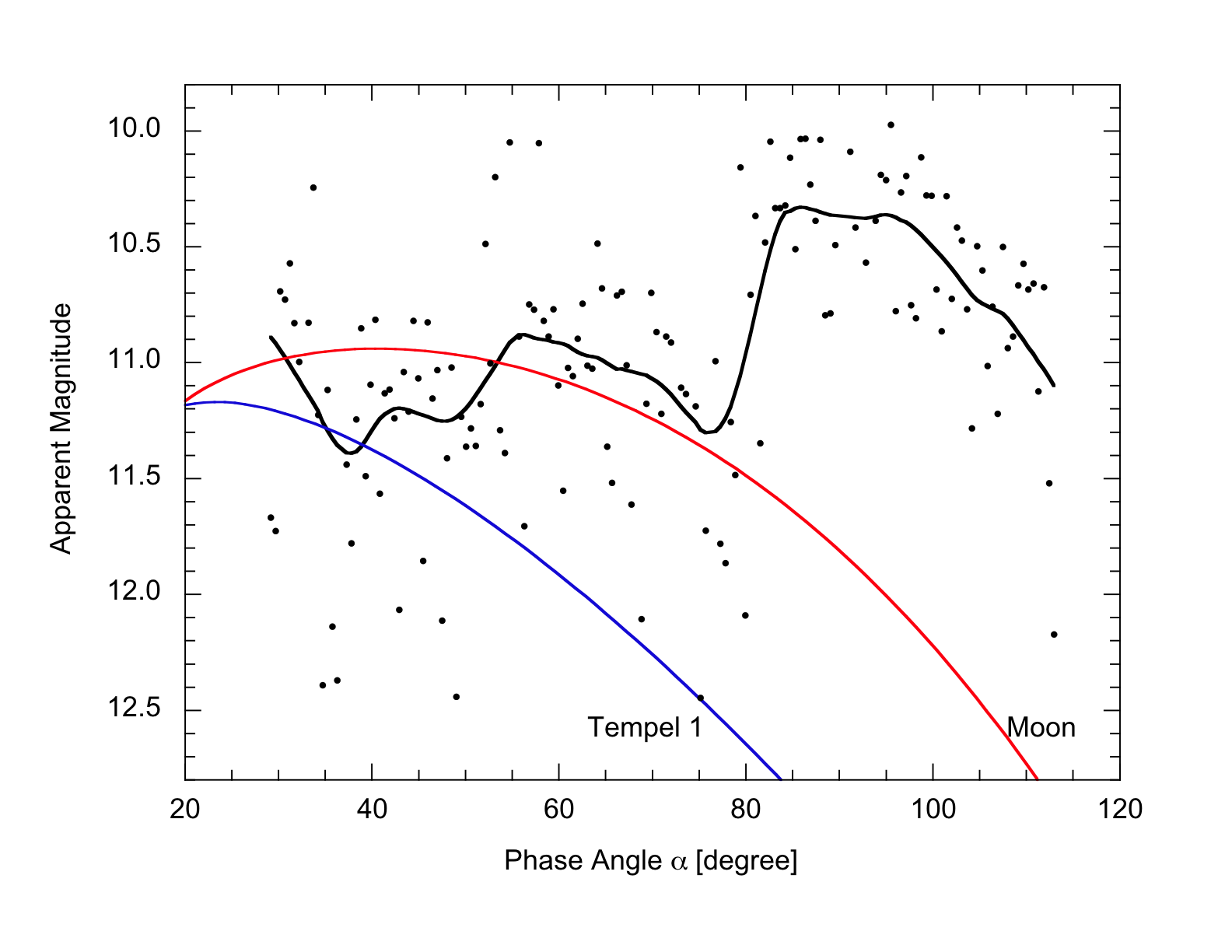}
\caption{Apparent magnitude vs. phase angle for Phaethon (black points).  The black line shows a running box mean capturing 15\% of the data.  Red and blue curves show the brightness variation of hypothetical bodies at the position of Phaethon and having phase functions as measured for the Moon (red) and the nucleus of comet P/Tempel 1 (blue). The latter two curves have been arbitrarily shifted in the vertical direction in order to facilitate easy comparison with the data.  \label{phasefunctions}} 
\end{center} 
\end{figure}

The brightening of Phaethon at $\alpha \sim$80$\degr$ occurs too suddenly to be caused by phase-angle dependent scattering effects.  For example, in Figure (\ref{phasefunctions}) we show phase curves for the Moon (Lane and Irvine 1973; red line) and for the nucleus of comet P/Tempel 1 (Li et al. 2007; blue line).  Both curves are corrected to the heliocentric and geocentric distances of Phaethon using the inverse square law so that they should provide a match to the Phaethon data if its surface is either Moon-like or comet-like.  Evidently, the Lunar phase function does match the data for $\alpha <$80$\degr$ but the agreement is very poor at larger angles, even given the wide scatter in the Phaethon data.  Other solid-body phase dependences, for example those shown by C-type and S-type asteroids (Bowell et al. 1989) were tested and likewise fail to match the Phaethon data.  Opposite to the asteroids, diffraction from dust particles with 2$\pi a/\lambda >$ 1 causes comets to be forward-scattering.  However, the cometary phase function also provides an unconvincing match to the Phaethon photometry because the forward-scattering peak begins at $\alpha >$ 100$^{o}$ and increases steadily up to $\alpha$ = 180$\degr$ (Kolokolova et al. 2004), whereas Phaethon's brightness decreases for $\alpha >$ 90$\degr$ (Figure \ref{phasefunctions}). For all these reasons, we discount the possibility that Phaethon's brightness variations are caused by simple phase-related effects.  

\begin{figure}[]
%\epsscale{0.95}
\begin{center}
%%\plotfiddle{f101.ps}{0.5cm}{-90.}{.4}{.4}{-10000}{.00}
%\plotone{figure6.pdf}
%\plotone{Ph_Moon_Ratio2.pdf}
%\plotone{jewitt_f7.pdf}
\includegraphics[width=1.0\textwidth]{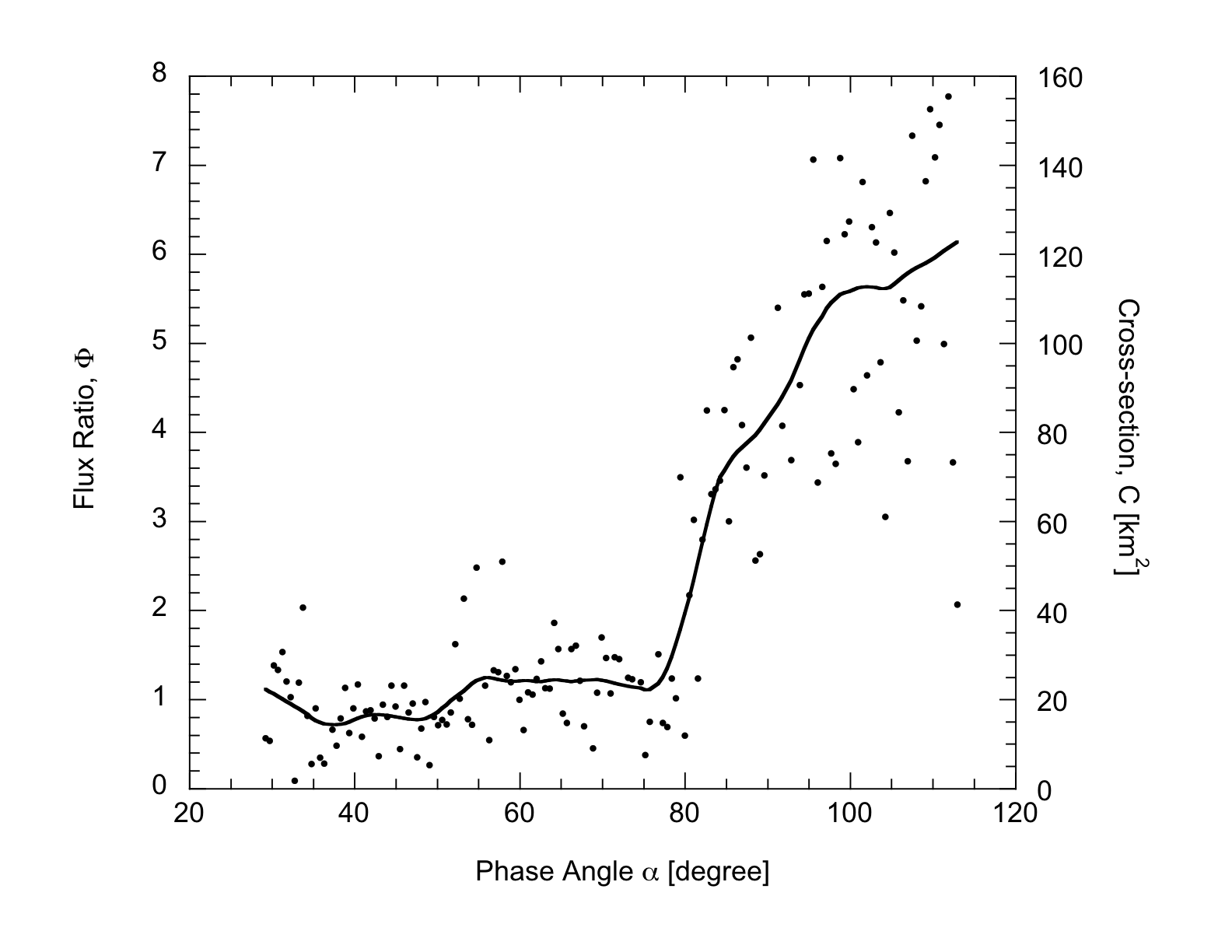}
\caption{Plot of $\Phi$, the ratio of the light from Phaethon to the light from the Moon as a function of phase angle, $\alpha$, normalized to $\Phi$ = 1 in the range 30$\degr$ $< \alpha <$ 80$\degr$.  The right-hand axis shows the effective cross-section in square kilometers, assuming that $\Phi$ = 1 corresponds to $C$ = 20 km$^2$. \label{Phi}} 
\end{center} 
\end{figure}

Nevertheless, given the large and changing phase angles of Phaethon (Table 1), it is appropriate to make an allowance for phase function effects in the interpretation of the photometry.  We note that the Lunar phase function provides a reasonable match to the data for $\alpha <$80$\degr$ (Figure \ref{phasefunctions}).  In Figure (\ref{Phi}) we show $\Phi$, the ratio of the signal from Phaethon to the Lunar phase curve, normalized such that the median value of $\Phi$ in the range 30$\degr$ $< \alpha <$ 80$\degr$ is unity.  The figure shows that the brightness of Phaethon grows dramatically relative to that of the Moon for larger phase angles.  For angles $\alpha >$ 110$\degr$ the noise in the ratio becomes excessive, since Phaethon appears very faint.  The right-hand axis in Figure (\ref{Phi}) shows the scattering cross-section, $C$ (km$^2$), computed from the brightness on the assumption that $\Phi$ = 1 corresponds to $C$ = $\pi r_e^2$ = 20 km$^2$.   This cross-section only has meaning if the phase function of the scatterers is equal to the phase function of the nucleus.  For example, if Phaethon ejected small, forward-scattering particles at $\alpha >$ 80$\degr$, then a smaller geometric cross-section would be indicated than is given by the right-hand axis in Figure (\ref{Phi}).  We possess too little information to place useful constraints on any change in the phase function caused by the outburst event.   Figure (\ref{Phi}) shows, however, that the brightening of Phaethon corresponds to an increase in the scattering cross-section at least comparable to, and perhaps many times larger than, the geometric cross-section of the bare nucleus of Phaethon.  For definiteness in the following discussion, we assume from Figure (\ref{Phi}) that the brightening of Phaethon corresponds to an increase in the effective cross-section by $C$ = 100 km$^2$, while accepting that this value is uncertain (because of the unknown phase function) by a factor of a few.

\section{Discussion}
\subsection{Solar Wind Interaction}
What is the origin of the brightening of Phaethon?  The probability that we are witness to dust ejected from Phaethon by recent impact is vanishingly small.  
The fact that the main brightening of Phaethon occurs soon ($\sim$1/2 day) after a brightening of the corona hints at a causal relationship, perhaps through solar wind excitation  of surface materials.   We first address this possibility using an energy argument.  The increase in brightness of Phaethon by 2 magnitudes (relative to the phase-darkened Lunar curve in Figure 6) corresponds to an increase in the photon flux density at Earth by $f_{\lambda}$ = 2$\times$10$^{-16}$ W m$^{-2}$ \AA$^{-1}$ (Drilling  and Landolt 2000).  Assuming isotropic emission, this corresponds to an emitted optical power $P_o$ = 4 $\pi \Delta^2 f_{\lambda} \Delta \lambda$ (W), where $\Delta$ is the geocentric distance and $\Delta \lambda$ = 1400 \AA~is effective width of the spectral response function of the camera.    The total energy of all the solar wind particles striking Phaethon, per second, is of order 
$P_w =  \pi r_e^2 \Delta V N_1 k T$, where $\Delta V$ is the speed of the wind sweeping past Phaethon, $N_1$ is the number density of particles in the wind, $k$ is the Stefan-Boltzmann constant and $T$ is the temperature of the particles in the wind.  We set $P_o = \zeta P_w$, where $\zeta$ is the efficiency with which solar wind energy can be converted into optical photons.  This gives 

\begin{equation}
N_1 = \frac{4 \Delta^2 f_{\lambda} \Delta \lambda}{r_e^2 \zeta \Delta V k T}
\end{equation} 

\noindent as the critical density needed for fluorescence to power the optical emission.  
We take $\Delta$ = 1 AU (Table 1), $\Delta V$ = 500 km s$^{-1}$, radius $r_e$ = 2.5 km, $\zeta$ = 1, $k$ = 1.38$\times$10$^{-23}$ J K$^{-1}$ and $T$ = 10$^6$ K to find $N_1 = $ 5$\times$10$^{14}$ m$^{-3}$.  The estimate is very crude in the sense that the emission is unlikely to be isotropic, emission outside the filter bandpass is neglected and all realistic processes have $\zeta <$ 1, tending to increase $N_1$ still further.  The required density is far higher than the typical densities in streamers when scaled to 0.14 AU ($N_1 \ll$10$^{10}$ m$^{-3}$, Li et al. 1998) or coronal mass ejections ($N_1 <$ 10$^{9}$ m$^{-3}$ (Reinard 2008)) so that we can dismiss the possibility that brightening is stimulated by charged particle impact.   By a similar argument we reject fluorescent emission caused by solar x-ray or ultraviolet photons.

\subsection{Dust}
We next consider the possibility that the brightening results from scattering from dust particles ejected from Phaethon and having a combined cross-section $C$ = 100 km$^2$, as noted above.  The mass corresponding to $C$ is dependent on the mean radius of the scatterers, $\overline a$, and given by $M \sim \rho \overline a C$ (kg).  Substituting $\rho$ = 2500 kg m$^{-3}$ and $C$ = 10$^8$ m$^2$ we find $M$ = 2.5$\times$10$^8 a_1$ kg, where $a_1$ is the particle radius expressed in millimeters.  For example, if $a_1$ = 1 (the nominal Geminid meteoroid size), the 2.5$\times$10$^8$ kg mass is equivalent to the loss of a monolayer from the surface.  The Geminid stream has a mass $M_s$ = 10$^{12}$ to 10$^{13}$ kg (Hughes and McBride 1989, Jenniskens 1994) and, therefore, the mass of dust implied by our photometry with $a_1$ = 1 is only 10$^{-4} M_s$.  (If the emitted particles are much smaller than 1 mm, their mass would be correspondingly reduced).  The  stream age is  $\sim$10$^3$ yr (Gustafson 1989, Ryabova 2007).  We conclude that mass loss events like the one observed cannot account for the Geminid stream mass unless they are frequent ($\sim$10 events per orbit).  Given the limited nature of published constraints on mass loss from Phaethon, it is difficult to rule out mass loss events with this frequency.  For example, mass loss could occur only close to perihelion, when ground-based observations are practically impossible, rendering all previously published limits irrelevant.

Sustained loss of particles from a small Solar system body requires both a source of particles and a means to eject them from the body.  Without a source, ejection processes will deplete the dust to a bare surface.  Without an ejection mechanism, the surface will clog with dust, impeding the production.  On comets, dust particles are exposed at the surface by the sublimation of entrapping ice, while drag forces from the sublimated gas  deplete dust grains by launching them to the interplanetary medium.  Could this process operate on Phaethon?

High surface temperatures are expected on the day-side of Phaethon as a result of its small perihelion distance.  The peak temperatures  cannot be exactly calculated  because they depend on many unknown or poorly-known quantities (including the obliquity and orientation of the nucleus spin vector relative to the line of apsides, and the thermal parameters of the surface).  We assess the approximate range of perihelion day-side temperatures as follows.  An isothermal, spherical blackbody in equilibrium with sunlight would have $T$ = 746 K at Phaethon's 0.14 AU perihelion distance. Since real objects sustain a day-night temperature contrast (i.e. they are not isothermal), this may be taken as an effective lower limit to the day-side temperature at perihelion.  An effective upper limit  is given by the subsolar temperature on a non-rotating body (or on one having a spin vector aligned with the Sun), which could reach $T$ = 1050 K at $q$ = 0.14 AU.  The peak day-side temperatures are thus contained in the range 746 $< T < $ 1050 K, in agreement with more complicated thermal models (Ohtsuka et al. 2009).  For comparison, day-side \textit{aphelion}  temperatures will be smaller by the factor ((1+e)/(1-e))$^{1/2} \sim$ 4, corresponding to 180 K and 256 K, for the low and high temperature limits, respectively.  Thus, any element of the surface of Phaethon must experience large temperature variations, $\delta T \sim$ 500 K, in the course of an orbit. Comparably large variations may be experienced even in the course of a nucleus rotation.  

Such high temperatures preclude the existence of water ice near the surface of Phaethon and thus eliminate comet-like water ice sublimation as a driving mechanism.  Could the activity instead be caused by the sublimation of ice buried at depth?  The timescale for the conduction of heat from the surface to the core of a spherical body of radius $r_e$ is roughly $\tau_c$ = $r_e^2/\kappa$, where $\kappa$ is the thermal diffusivity of the material.  Common dielectrics have $\kappa \sim$10$^{-6}$ m$^{2}$ s$^{-1}$.  Substituting $r_e$ = 2.5 km, we obtain $\tau_c$ = 2$\times$10$^5$ yr.  The dynamical lifetime of Phaethon in its present orbit is unknown and can only be guessed statistically; the median dynamical lifetime of bodies moving in similar orbits is $\tau_d \sim$ 26 Myr (de Leon et al. 2010).  With $\tau_d \gg \tau_c$, we can assume that heat deposited on the surface of Phaethon has conducted to the deep interior.    The core temperature calculated as in Jewitt and Hsieh (2006), 
$T_c$ = 300 K, is too high for water ice to survive in the core.  Accordingly, we consider it unlikely that on-going activity in Phaethon could be driven by the sublimation of deeply buried water ice; Phaethon is simply too hot, even in its core.  A separate argument against deeply-rooted activity of any kind is the observation that the activity in Phaethon coincides with perihelion.  This coincidence can only be explained if the source of the activity lies less than a few thermal skin depths (a few $\times$0.1-m) beneath the physical surface, otherwise the activity source would be thermally (and temporally) decoupled from the heat of the Sun.

\subsection{Dust Production: Thermal Decomposition}

While ice sublimation is unlikely, other heat-triggered processes might operate at the extreme perihelion temperatures on Phaethon. The surface is not hot enough for rock itself to significantly sublimate, but the perihelion temperatures exceed those needed to thermally decompose some rocks. For example, experiments with the hydrated mineral serpentine show structural changes on laboratory timescales (hours and days) beginning at $T \sim$ 600 K (Akai 1992, Nozaki et al. 2006, Nakato et al. 2008). These changes become extensive by 900 K, comparable to the peak temperatures anticipated on Phaethon.  Chemically bound water is progressively lost as the temperature rises, and there are associated changes in the crystal structure and general shrinkage of the rock leading to internal stress and cracking.  The loss of water from dehydration could provide a source of gas, although it is not obvious that gas drag forces produced this way would be sufficient to eject solid particles against nucleus gravity.  Dehydration of phyllosilicates will result in shrinkage and cracking, with associated dust production, just as occurs in sun-baked mud-flats on Earth.   

Based on a numerical model, Bottke et al. (2002) found that Phaethon has a 20\% likelihood of originating in the central main-belt (2.5 $< a <$ 2.8 AU) and an 80\% likelihood of originating in the inner main-belt ($a <$ 2.5 AU).  A specific association with asteroid (2) Pallas ($a$ =  2.77 AU) has been recently proposed (de Leon et al. 2010).  Asteroids in this region of the belt include a large proportion of volatile-rich $C$-types that may contain phyllosilicates (clays) or other hydrated minerals. Indeed, these materials have been specifically suggested as components of Phaethon (Licandro et al. 2007, Ohtsuka et al. 2009).  Thus, there exist good reasons to think that the surface of Phaethon might be susceptible to the high perihelion temperatures by cracking and the production of dust through thermal decomposition.

\subsection{Dust Production: Thermal Fracture}

Another  potential producer of particles is cracking resulting from  stresses induced by differential thermal expansion (Oliva et al. 1971, Lauriello 1974).  It is well known from everyday experience that thermal fracture can be induced by strong temperature gradients (imposed by heating a body on a timescale shorter than its thermal conduction time, as when hot water is poured into, and cracks, a cold glass).  On Phaethon, strong diurnal temperature variations ($\delta T \sim$ 500 K) are expected in association with the $\sim$3.6 hr rotation period (Krugly et al. 2002). More generally, thermal fracture occurs even when the  temperature changes very slowly, because rocks have a granular structure and the expansivities of grain and inter-grain material are typically different.   Thermal cracking has been studied experimentally using sound waves produced when cracks open around mineral grains in heated rocks (Chen and Wang 1980). Experiments in the laboratory show that thermal fracture proceeds down to the grain size in rocks.  In Terrestrial rocks and meteorites, the natural grain size is commonly near 1-mm, providing an approximate but intriguing match to the characteristic sizes of Geminid meteoroids.

In the elastic regime, the stress on a material caused by thermal expansion 
is just

\begin{equation}
S =  \alpha Y \delta T 
\label{stress}
\end{equation}

\noindent where $Y$ is Young's Modulus, $\alpha$ is the thermal expansivity and $\alpha \delta T$ is the strain resulting from thermal expansion from a temperature change $\delta T$.  Values $Y$ = (10 to 100)$\times$10$^9$ N m$^{-2}$ are typical for rock (Pariseau 2006), while the thermal expansivities of common rocks are within a factor of a few of $\alpha$ = 10$^{-5}$ K$^{-1}$ (Lauriello 1974, Richter and Simmons 1974), which we take as our best-guess estimate of the expansivity of Phaethon.  Substituting $\delta T$ = 500 K (from the previous section) in Equation (\ref{stress}) we obtain thermal stress $S$ = (5 to 50)$\times$10$^7$ N m$^{-2}$ (500 to 5000 bars) corresponding to the temperature changes experienced on Phaethon as it moves around its orbit. Depending on the thermal diffusivity, comparable temperature variations and thermal stresses may also be experienced on a much shorter timescale as Phaethon rotates.  The thermal stress is larger than the $\sim$100 bar yield strength characteristic of rocks in tension (Lauriello 1974).  Thus, thermal cracking should be expected and could be an important mechanism for producing small particles on Phaethon.

The highest temperatures and thermal stresses will be concentrated near the surface, in a layer having thickness comparable to the diurnal thermal skin depth, $d \sim (\kappa P)^{1/2}$, where $\kappa $ (m$^2$ s$^{-1}$) is the thermal diffusivity and $P$ the rotation period of Phaethon.  Substituting $\kappa$ = 10$^{-6}$ m$^2$ s$^{-1}$  and $P$ = 3.6 hrs (Krugly et al. 2002), we estimate $d$ = 0.1 m.   Representing Phaethon as a sphere with effective radius $r_e$ = 2.5 km, the mass of material accessible to thermal fracture at any time is $\Delta M$ = $4 \pi r_e^2  \rho d$, where $\rho$ is the density, here taken to be $\rho$ = 2500 kg m$^{-3}$.  Substituting, we find that strong thermal gradients can affect $\Delta M \sim$ 10$^{10}$ kg of material on Phaethon.  This is larger than the inferred dust mass but two to three orders of magnitude too small to be consistent with the Geminid stream mass. However, thermal fracture will continue so long as fresh material is exposed and the resulting debris cleared away. As with dust produced by dehydration of hydrated minerals, the rate of progression depends on how quickly particulate material on Phaethon can be ejected against self-gravity.  

\subsection{Dust Loss Mechanisms}
As noted above, many dust production mechanisms become self-limiting if there is no process by which freshly produced particles can be removed from the surface.   Here, we briefly consider three processes that might, singly or in combination, lead to the escape of dust from Phaethon.  

Thermal fracture itself is expected to launch freshly produced particles with a non-zero velocity.   To see this, we consider a flat plate of cross-section $A$ and thickness $\ell$, with a temperature difference impressed between the two sides of the plate, $\delta T$.  The thermal expansion distance between the two flat sides is $\delta \ell$ = $\alpha \ell \delta T$.  The force exerted by the expansion is $F$ = $Y (\delta \ell / \ell) A$.  The approximate work done by the expansion is the product of these two quantities, $W \sim Y (\delta \ell/\ell) ( \delta \ell) A$.  Substituting, we obtain $W$ = $Y \alpha^2 \delta T^2 \ell A$.  Suppose that some fraction, $\eta$, of this thermal strain energy can be converted into kinetic energy of the fractured material, $E \sim \rho \ell A V^2$.  Setting $E = \eta W$, obtain

\begin{equation}
V \sim \alpha \delta T \sqrt{\frac{\eta Y}{\rho}}
\label{fracture}
\end{equation}

\noindent as a crude estimate of the characteristic speeds of particles produced by fracture.  We adopt the same values for $\alpha$, $\delta T$, $Y$ and $\rho$ as in Section 3.4.  The magnitude of the energy conversion efficiency, $\eta$, is unknown and reflects a competition between energy used to fracture the material, energy radiated away as vibrations, and thermodynamic and other effects.  We assume $\eta$ = 1 to give an upper bound to the particle velocity.  Then, Equation (\ref{fracture}) gives $V \sim$ 10 to 30 m s$^{-1}$ as the characteristic particle speed.  This is an order of magnitude larger than the gravitational escape speed from Phaethon.  Therefore, thermal fracture should be capable of launching particles above the escape speed if the efficiencies $\eta >$ a few percent.  While $\eta$ is unknown, it would not be surprising to find that a significant fraction of the particles produced by thermal fracture can be launched directly from the surface into interplanetary space.
 
Electrostatic forces offer another mechanism to remove dust.  Evidence from the Moon shows that dust particles are periodically  lifted from the surface at speeds $\sim$1 m s$^{-1}$ by steep electric field gradients that develop in the terminator regions (de and Criswell 1977, Colwell et al. 2007).  Strong electric field gradients should be present on any low-conductivity surfaces exposed to the Solar radiation and the Solar wind. Indeed, electrostatic levitation has been implicated in, for example, mobilizing dust and forming ``ponds'' on 16 km mean diameter asteroid (433) Eros (Colwell et al. 2005).  Even neglecting centripetal effects from its rapid rotation, the gravity on Phaethon is $\sim$3 times smaller than on Eros and $\sim$700 times smaller than on the Moon; the dynamical effects of electrostatic charging should be proportionally greater.  Dust speeds sufficient to produce only ballistic jumps on the Moon may be sufficient to produce escape from the gravitational field of a body as small as Phaethon.  The small heliocentric distance of Phaethon at perihelion will not enhance the magnitude of the electrostatic forces (which depend ultimately on the photo-electric effect, not the distance) but the photocurrent rates should be $\sim$50 times larger than on the Moon because of Phaethon's smaller distance.   

Finally, the small perihelion distance of Phaethon can help to sweep dust particles from its surface through the action of radiation pressure.  
To see this, we assume that the nucleus can be approximated by an ellipsoid with axes $a > b = c$, in rotation about the $c$ axis.  Measurements of the rotational lightcurve define the effective cross-section, $\pi r_e^2$, and set a limit to the ratio of the equatorial axes, $f = a/b$.  The data give  $r_e$ = 2.5 km and $f$ = 1.4.  Setting $\pi r_e^2$ = $\pi a b$ we may write $a = r_e f^{1/2}$ and $b = r_e f^{-1/2}$. The volume of the ellipsoid, $V = 4/3 \pi a b^2$ can then be written in terms of the two measurable quantities $r_e$ and $f$ as $V = 4/3 \pi r_e^3 f^{-1/2}$. Rotation about the $c$-axis with period, $P$, gives rise to a centripetal acceleration which is largest at the apex of the figure. There, the net acceleration towards the center may be written

\begin{equation}
g = \frac{4 \pi G \rho r_e}{3 f^{3/2}} - \frac{4 \pi^2 r_e f^{1/2}}{P^2}
\label{gravity}
\end{equation}

\noindent where the first term is the gravitational attraction and the second is the centripetal acceleration.  Equation (\ref{gravity}) may be compared with the acceleration due to radiation pressure, 

\begin{equation}
g_{pr} = \beta g_{\odot} / R_{au}^2, 
\label{radp}
\end{equation}

\noindent where $g_{\odot}$ = 0.006 m s$^{-2}$ is the gravitational acceleration to the Sun at 1 AU, and $R_{au}$ is the heliocentric distance expressed in AU. The dimensionless quantity $\beta$ is principally a function of particle size, $a$, but also depends upon the grain shape, composition and porosity.  As a useful first approximation, we take $\beta \sim 1/a$, where $a$ is expressed in microns.  Comparing Equation (\ref{gravity}) to Equation (\ref{radp}) gives the critical value of  radiation pressure efficiency,  $ \beta_c = a_c^{-1}$, above which a particle will be unbound to the nucleus.  We obtain

\begin{equation}
\frac{1}{a_c} =  \frac{4 \pi R_{au}^2 f^{1/2} r_e}{3 g_{\odot}}\left[\frac{G \rho}{f^2} - \frac{3 \pi}{P^2}\right]
\label{betac}
\end{equation}

\noindent in which $a_c$ is expressed in microns.  

\begin{figure}[]
%\epsscale{0.95}
\begin{center}
%%\plotfiddle{f101.ps}{0.5cm}{-90.}{.4}{.4}{-10000}{.00}
%\plotone{figure6.pdf}
%\plotone{ac_vs_q_plot.pdf}
%\plotone{jewitt_f8.pdf}
\includegraphics[width=1.0\textwidth]{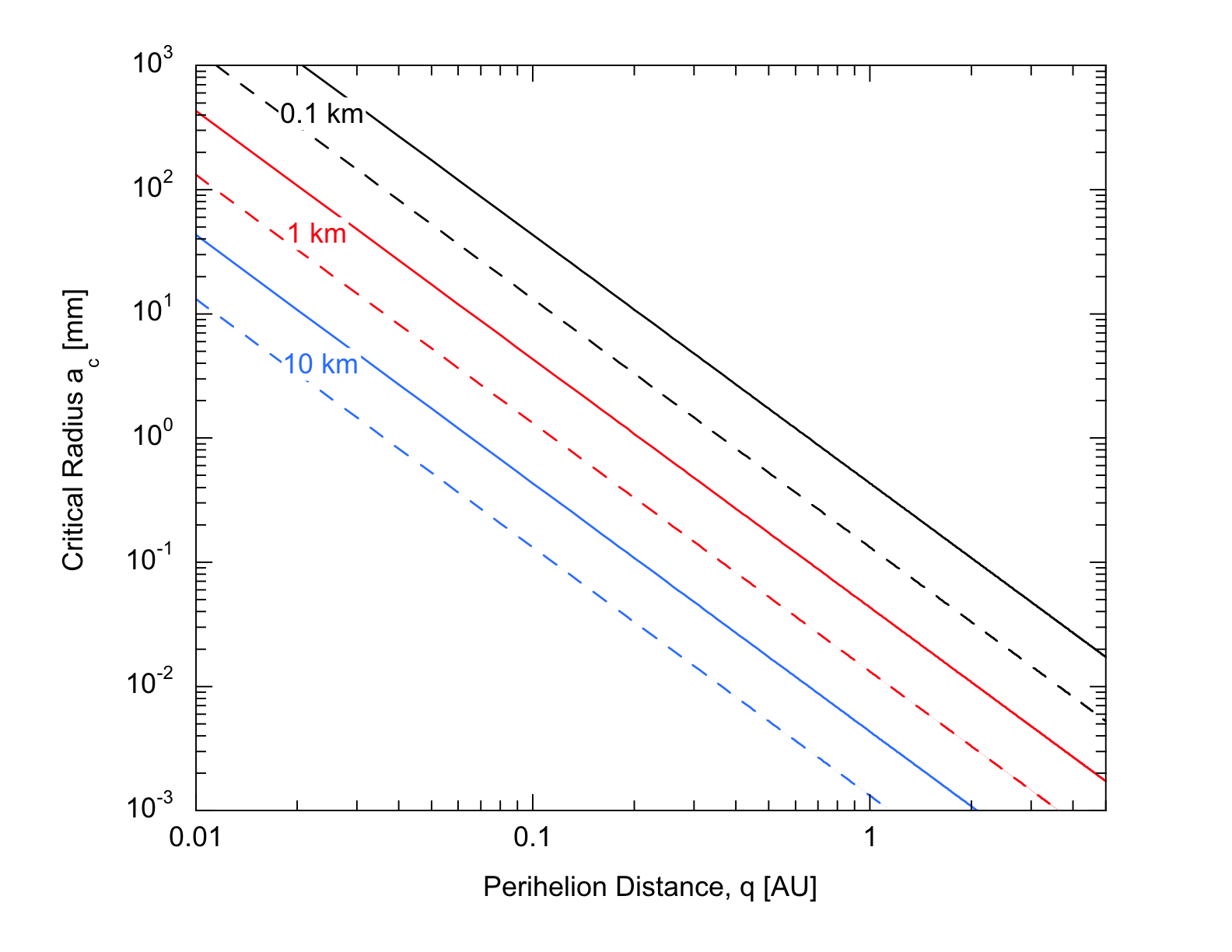}
\caption{Critical dust radius for radiation pressure sweeping as a function of perihelion distance.  Black, red and blue lines correspond to asteroid radii $r_e$ = 0.1, 1 and 10 km, respectively.  Solid lines are for body axis ratios $f$ = 1.4, dashed lines are for spherical bodies, $f$ = 1.  All other parameters in Equation (\ref{betac}) are the same as those for Phaethon. \label{ac_vs_q}} 
\end{center} 
\end{figure}

Equation (\ref{betac}) is presented graphically in Figure (\ref{ac_vs_q}).  Substituting the nominal values $R_{au}$ = 0.14, $f$ = 1.4, $r_e$ = 2.5 km, $\rho$ = 2500 kg m$^{-3}$ and $P$ = 3.6 hrs into Equation (\ref{betac}) we obtain $a_c$ = 10$^{3}~\mu$m = 1 mm.  Particles smaller than $a_c$ experience radiation pressure forces larger than their weight and can be stripped from the nucleus if they are briefly detached from the surface and the net force vector acting upon them aims away from the nucleus.  The latter condition is naturally satisfied at the terminator, where electric field gradient forces are the largest (Colwell et al. 2007).  Equation (\ref{betac}) is approximate in the sense that we have assumed a prolate nucleus ($b$ = $c$) whereas a triaxial approximation may be more accurate.  We have also assumed the density, $\rho$, and that $f$ measured from the lightcurves represents the true axis ratio whereas in fact it is only a lower limit to the ratio because of the effects of projection.   Given these many approximations and unknowns, we cannot prove that radiation pressure sweeping, in combination with electrostatic ejection and the expected fracture launch speeds, is responsible for cleaning particles from Phaethon's surface.  However, these effects combined do constitute  a plausible path by which particles can be removed from Phaethon.  The small size, $r_e$, small perihelion distance, $R_{au}$, and short rotation period, $P$, all maximize $a_c$ relative to larger, more distant and more slowly rotating asteroids.

To conclude, we speculate that dust production and loss rates are both high as consequences of the very high surface temperatures on Phaethon at perihelion. The comet-like activity does not imply that sublimating ice is responsible in this object. Instead, it is appropriate to think of Phaethon as a rock comet in which thermal fracture, dehydration cracking, radiation pressure sweeping and electrostatic effects may all play roles in producing and removing particles from the surface.  Several of these processes decline precipitously with increasing distance from the Sun.  For example, if displaced from $q$ = 0.14 AU to the main-belt (3 AU), thermal fracture would become unimportant and the critical dust size for radiation pressure sweeping on Phaethon would decrease by nearly a factor of 10$^3$, from $\sim$1 mm to $\sim$1 $\mu$m (Figure \ref{ac_vs_q}).  Of the 19 known asteroids having perihelia smaller than Phaethon's, few have been searched for evidence of mass loss or, indeed, subjected to any meaningful physical study. An important observational next-step will be to examine these objects near perihelion in search of comet-like activity analogous to that observed in Phaethon.  Unfortunately,  none is bright enough to be observed using STEREO, so that relevant observations will be difficult to obtain.  It will also be important to examine Phaethon at future perihelion passages, to determine the frequency of mass-loss events and so to decide whether the meteoroid stream is in steady-state.

\clearpage

\section{Summary}
We undertook a careful photometric analysis of asteroid (3200) Phaethon, the purported Geminid meteoroid parent, when near perihelion (0.14 AU) in June 2009.  We used images from NASA's STEREO-A coronal-monitoring spacecraft, to obtain the following results:

\begin{enumerate}
\item The apparent near-perihelion brightness doubled unexpectedly starting UT 2009 June 20.2$\pm$0.2, with the transition from faint state to bright state occurring over only $\sim$0.1 day.  The morphology remained point-like in images having pixel size 70$\times$70 arcseconds (51,000$\times$51,000 km at Phaethon).

\item The brightening is of the wrong sign and too sudden to be attributed to the phase function of Phaethon or to other geometric effects, and too large to be due to plasma impact excitation or fluorescence of the surface. The brightening is best interpreted as due to the ejection of dust particles having a combined cross-section comparable to, or larger than, Phaethon's.

\item If the ejected particles are of millimeter size, their combined mass must be $\sim$2.5$\times$10$^8$ kg, or about 10$^{-4}$ times the mass of the Geminid stream.   Given the 10$^3$ yr stream age, about 10 comparable events per orbit would be needed to replenish the stream mass, in steady-state.  

\item Normal cometary mass-loss driven by ice sublimation is unlikely on Phaethon, since the surface and the interior are too hot to retain ice on timescales comparable to the dynamical lifetime. 

\item Instead, we interpret Phaethon as a ``rock comet'', in which dust is produced by the thermal decomposition and cracking of hydrated minerals, and by thermal fracture at the high surface temperatures ($\sim$1000 K) experienced when close to the sun.  Near perihelion, particles smaller than about 1 mm in radius cannot be held by Phaethon against radiation pressure, which thus acts as a cleaning agent for this body.

\end{enumerate}

\acknowledgements
We thank Abby Kavner, Jane Luu and Ed Young for pointers to the literature concerning thermal fracture and dehydration, Michal Drahus, Aurelie Guilbert, Toshi Kasuga, Pedro Lacerda and Bin Yang for comments on the manuscript and the referee for a speedy review.
The Heliospheric Imager (HI) instrument was developed by a collaboration which
included the University of Birmingham and the Rutherford Appleton Laboratory,
both in the UK, and the Centre Spatial de Liege (CSL), Belgium, and the US
Naval Research Laboratory (NRL), Washington DC, USA.   The CME catalog is 
generated and maintained at the CDAW Data Center by NASA and The Catholic 
University of America in cooperation with the Naval Research Laboratory. 
SOHO is a project of international cooperation between ESA and NASA. This 
work was supported, in part, by a grant from
NASA's Planetary Astronomy Program.

\begin{deluxetable}{llllcl}
\tabletypesize{\scriptsize}
\tablecaption{Observational Geometry for HI-1A Data
\label{geometry}}
\tablewidth{0pt}
\tablehead{
\colhead{UT Date} & \colhead{DOY\tablenotemark{a}} & \colhead{$R$ [AU]\tablenotemark{b}} & \colhead{$\Delta$ [AU] \tablenotemark{c}}   & \colhead{$\alpha$ [deg]\tablenotemark{d}} 
& \colhead{$\epsilon$[deg]\tablenotemark{e}}   }
\startdata
2009-Jun-17 &  168 	& 0.197 &	1.131 &	24.7 & 4.9  \\
2009-Jun-18 &	 169 	& 0.171 &	1.090 &	35.7 & 6.0  \\
2009-Jun-19 &	 170 	& 0.151 &	1.042 &	51.9 & 7.1  \\
2009-Jun-20 &	 171 	& 0.141 &	0.989 &	72.8  & 8.1  \\
2009-Jun-21 &	 172 	& 0.143 &	0.932 &	95.6 & 8.6  \\
2009-Jun-22 &	 173 	& 0.158 &	0.877 &	115.8 & 8.5  \\

\enddata

%% Text for table notes should follow after the \enddata but before
%% the \end{deluxetable}. Make sure there is at least one \tablenotemark
%% in the table for each \tablenotetext.

\tablenotetext{a}{Day of year.}
\tablenotetext{b}{Heliocentric distance in AU.}
\tablenotetext{c}{Object to spacecraft distance in AU.}
\tablenotetext{d}{Phase angle (Sun-Phaethon-spacecraft) [degrees].}
\tablenotetext{e}{Elongation angle (Sun-spacecraft-Phaethon) [degrees]}

\end{deluxetable}

\clearpage

\begin{deluxetable}{llllll}
\tabletypesize{\scriptsize}
\tablecaption{Averaged Photometry
\label{nucleus}}
\tablewidth{0pt}
\tablehead{
\colhead{Phase angle\tablenotemark{a}} & \colhead{$\overline{R}$\tablenotemark{b}} & \colhead{$\overline{\Delta}$ \tablenotemark{c}} & \colhead{$N$\tablenotemark{d}} & \colhead{$m_R$\tablenotemark{e}}    }
\startdata
30  $\le \alpha <$ 40 &	0.172 & 	1.094 & 19 & 11.35$\pm$0.19 \\
40  $\le \alpha <$ 50 &	0.157 & 	1.061 & 19 & 11.32$\pm$0.11 \\
50  $\le \alpha <$  60 &	0.148 & 	1.034 & 19 & 10.91$\pm$0.11 \\
60  $\le \alpha <$  70 &	0.143 & 	1.008 & 18 & 11.08$\pm$0.10 \\
70  $\le \alpha <$  80 &	0.140 & 	0.982 & 16 & 11.32$\pm$0.14 \\
80  $\le \alpha <$  90 &	0.140 & 	0.958 & 18 & 10.41$\pm$0.08 \\
90  $\le \alpha <$  100 &	0.143 & 	0.933 & 15 & 10.35$\pm$0.07 \\
100  $\le \alpha <$  110 &	0.148 & 	0.907 & 19 & 10.73$\pm$0.06 \\
\enddata

%% Text for table notes should follow after the \enddata but before
%% the \end{deluxetable}. Make sure there is at least one \tablenotemark
%% in the table for each \tablenotetext.

\tablenotetext{a}{Phase angle range over which photometry was averaged}
\tablenotetext{b}{Average heliocentric distance, AU}
\tablenotetext{c}{Average geocentric distance, AU}
\tablenotetext{d}{Number of measurements used to compute the average}
\tablenotetext{e}{Average apparent magnitude and standard deviation}

\end{deluxetable}

\clearpage

\clearpage

\clearpage

\clearpage

\clearpage

\clearpage

\clearpage

\clearpage

\end{document}